\documentclass[%
 reprint,
superscriptaddress,
nofootinbib,
 amsmath,amssymb,showpacs,
 aps,
prl,
]{revtex4-1}

\usepackage{graphicx}
\usepackage{dcolumn}
\usepackage{bm}
\usepackage{subfig}
\usepackage{xfrac}
\usepackage{amsmath}
\usepackage{hyperref}
\usepackage{multirow}

\usepackage{color}

\DeclareCaptionJustification{justified}{\leftskip=0pt \rightskip=0pt \parfillskip=0pt plus 1fil}
\newcommand{\beqar}{\begin{eqnarray}}
\newcommand{\eeqar}{\end{eqnarray}}

\newcommand{\bcen}{\begin{center}}
\newcommand{\ecen}{\end{center}}

\newcommand{\beq}{\begin{equation}}
\newcommand{\eeq}{\end{equation}}
\newcommand{\beqa}{\begin{eqnarray}}
\newcommand{\eeqa}{\end{eqnarray}}

\begin{document}

\title{Invariant-based inverse engineering of time-dependent, coupled harmonic oscillators}

\author{A. Tobalina}
\email{ander.tobalina@ehu.eus}
\affiliation{Department of Physical Chemistry, University of the Basque Country UPV/EHU, Apdo 644, Bilbao, Spain}
\author{E. Torrontegui}
\affiliation{Instituto de F\'\i sica Fundamental IFF-CSIC, Calle Serrano 113b, 28006 Madrid, Spain}
\author{I. Lizuain}
\affiliation{Department of Applied Mathematics, University of the Basque Country UPV/EHU, Donostia-San Sebastian, Spain}
\author{M. Palmero}
\affiliation{Science and Math Cluster, Singapore University of Technology and Design, 8 Somapah Road, 487372 Singapore}
\author{J. G. Muga}%
\affiliation{Department of Physical Chemistry, University of the Basque Country UPV/EHU, Apdo 644, Bilbao, Spain}

\begin{abstract}
Two-dimensional systems with time-dependent controls   admit a quadratic Hamiltonian modelling near potential minima. 
Independent, dynamical normal modes 
facilitate inverse Hamiltonian engineering to control the system dynamics, but  
some systems are not separable into independent modes by a point transformation.     
For these ``coupled systems''   
2D invariants may still guide the Hamiltonian design. The theory to perform the inversion and  
two application examples are provided: (i) We control the deflection of wave packets in transversally harmonic waveguides; and (ii) we design the   state transfer  from one coupled oscillator to another.    
\end{abstract}

\maketitle

{\it Introduction.}
\label{sec:intro}
Controlling the 
motional dynamics of quantum systems is of paramount importance for 
fundamental science and quantum-based 
technologies \cite{Kielpinski2002}.  
Often the external driving needs to be fast,
but also gentle, to avoid   
excitations.
Slow adiabatic driving is gentle in this  sense, but it exposes the system 
for long times to control noise, heating, and perturbations. 
Shortcuts to adiabaticity (STA) are techniques to reach, 
via fast non-adiabatic routes, the results of slow adiabatic processes \cite{Torrontegui2013,Guery2019}. 
A distinction can be made between: STA methods  that keep the structure of some 
Hamiltonian form and design the time dependence of the controls, e.g. using invariants \cite{Chen2010_063002};  and those techniques that add new terms, e.g.  counterdiabatic driving \cite{Demirplak2003}.
Both  may be useful depending on system-dependent practical considerations. A frequent problem with 
added terms is the difficulty to implement them, whereas a limitation of structure-preserving, invariant-based methods is that 
they need  Hamiltonian-invariant pairs with specific  forms, such as  the Lewis-Leach family 
of Hamiltonian-invariant pairs \cite{Lewis1982}, to go beyond brute-force 
parameter optimization \cite{Torrontegui2013,Guery2019}. 

The  eigenvectors of
Lewis-Riesenfeld ``time-dependent invariants'' 
 \cite{Lewis1969}, with appropriate phase factors, 
are independent solutions of the Schr\"odinger equation and span a basis to expand any solution with constant expansion coefficients.
These invariants are useful to inverse engineer the Hamiltonian and  drive some desired dynamics 
 \cite{Chen2010_063002}. The multiplicity of solutions for the  trajectories of the control parameters,  allows for adjustments or optimization with respect to different objectives or cost functions \cite{Ruschhaupt2012}. The multiplicity is also very helpful when several oscillators have to be controlled simultaneously \cite{Palmero2014,Palmero2015a}.      

This work 
extends the domain of systems  that can be controlled by invariant-based inverse engineering. 
We shall deal with two-dimensional (2D) systems with quadratic Hamiltonians, found in particular in  
small-oscillation regimes of ultracold atom physics.
In fact quadratic Hamiltonians are ubiquitous as they represent the systems near potential minima \cite{Urzua2019}. For time-independent Hamiltonians the dynamics is simple to  
describe and, possibly, manipulate by finding  normal modes for effective uncoupled oscillators. This decomposition though, may not
be possible if the Hamiltonian parameters depend on time.       
Lizuain et al. \cite{Lizuain2017} described the condition for which a point transformation of coordinates decouples the instantaneous modes leading to truly independent ``dynamical normal modes'' \cite{Palmero2014} for two time-dependent harmonic oscillators: the  principal axes of the potential  should not rotate in the 2D space. 

When  the two dynamical-mode motions separate, inverse engineering the dynamics to perform some fast 
operation free from final excitations is relatively easy: each of the time-dependent effective oscillators implies a one-dimensional Hamiltonian-invariant ``Lewis-Leach'' pair \cite{Lewis1982} for which inverse engineering can be performed. The two oscillators have to be driven simultaneously with common controls  but,
among the plethora of  parameter trajectories, it is 
possible to find the ones that satisfy simultaneously the 
boundary conditions imposed on both oscillators. This strategy 
has been successfully applied to design the driving of  different operations on two trapped ions such as transport or expansions 
\cite{Palmero2014,Palmero2015a}, separation of two equal ions in double wells \cite{Palmero2015}, phase gates
\cite{Palmero2017}, or dynamical exchange cooling  \cite{Sagesser2020}.  

If the effective potential rotates, the motions do not separate, so inverse engineering the external driving 
cannot in principle be done using two independent 1D Hamiltonian-invariant pairs. Solutions to the ensuing control problem exist that depend on the system and/or the operation, such as taking refuge in a perturbative regime \cite{Palmero2017}, adding terms to cancel 
the inertial effects \cite{Lizuain2017}, increasing the number of time-dependent controls to uncouple the modes \cite{Sagesser2020}, 
or using more complex, non-point transformations to find independent modes \cite{Lizuain2019a}.  Here we explore instead 
the use of 2D dynamical invariants associated with the coupled Hamiltonian. 

{\it Hamiltonian model.\label{model}}
Consider the Hamiltonian 
\begin{equation}
    H(t) = \frac{ {p}_1^2}{2 } + \frac{ {p}_2^2}{2}
+\frac 1 2  \omega_1^2(t)  {q}_1^2 + \frac 1 2  \omega_2^2(t)  {q}_2^2 - {\gamma}(t)  {q}_1  {q}_2.
    \label{hamiltonianmw}
\end{equation}
We use throughout dimensionless variables such that no mass factors or $\hbar$ appear explicitly.    
Eq. (\ref{hamiltonianmw}) describes  
different physical systems, such as a single particle in a 2D potential, or two coupled harmonic oscillators on a line.  Other systems different from (one or two) particles 
but driven by Hamiltonians of the { form} (\ref{hamiltonianmw}) are, e.g.,  coupled superconducting qubits \cite{Barends2013,Rol2019, Peropadre2013, Garcia-Ripoll2020, Chen2014} or optomechanical oscillators  \cite{Kleckner_2011,Zhang2014, Aspelmeyer_2014}.  
All these systems are analogous to each other but, arguably, the single particle 
 in a 2D potential  is easiest to visualize
 so we shall use  a terminology (such as longitudinal and  transversal 
 directions for principal axes, rotations...) borrowed from that system. Indeed, our first example, see below,  deals 
 with a single particle.

The Hamiltonian (\ref{hamiltonianmw}) may be instantaneously diagonalized by  ``rotated''  variables
\cite{Lizuain2017} 
\beq
\label{variablechange}
\begin{pmatrix}
q_l \\ 
q_t
\end{pmatrix}
= A (t)
\begin{pmatrix}
q_1 \\ 
q_2 
\end{pmatrix}
,
\hspace{0.6cm}
\begin{pmatrix}
p_l \\ 
p_t
\end{pmatrix}
= A(t)
\begin{pmatrix}
p_1 \\ 
p_2 \\
\end{pmatrix},
\eeq
%
where 
{\small $
{A(t)=
\begin{pmatrix}
 \cos \theta(t) &  \sin \theta(t) \\
-\sin \theta(t) &  \cos \theta(t) \\
\end{pmatrix}}
$}, and 
%
%
\beq
\theta(t)=\frac 1 2 \arctan\!{\left(\!\frac{2 \gamma(t)}{\omega_2^2(t)-\omega_1^2(t)}\!\right)}.
\eeq
Subscripts $l$ and $t$ stand for ``longitudinal'' and ``transversal''. 
The original Hamiltonian, expressed in terms of the new variables, is  
\begin{eqnarray}
    H &=& \frac{ p_l^2}{2} + \frac{ p_t^2}{2} + \frac 1 2 \Omega_l^2  q_l^2 + \frac 1 2  \Omega_t^2  q_t^2,
     \label{insthamiltonian}
     \\
\label{nmfreq}
\Omega_l^2 &=& {\left(\omega_1^2+\omega_2^2-\Lambda\right)}/{2},\,\,\,\,\,\,
\Omega_t^2={\left(\omega_1^2+\omega_2^2+\Lambda\right)}/{2},
\end{eqnarray}
where $\Lambda(t)=\sqrt{4 \gamma^2(t) + [\omega_2^2(t)-\omega_1^2(t)]^2}$.

The formal decoupling in Eq. (\ref{insthamiltonian}) is a mirage. $H$ is not the Hamiltonian that describes the dynamics in the rotated variables $\{p_l, p_t, q_l, q_t\}$ \cite{Goldstein2002,Lizuain2017}. In general the dependence of $A(t)$ on time couples dynamically the ``instantaneous normal modes'', i.e., the normal modes that would  separate  the motion if the Hamiltonian 
kept for all times the values that the parameters have at a particular instant.  
In the moving frame the oscillators are coupled by a term proportional to $\dot{\theta}=d\theta/dt$ \cite{Lizuain2017}.    
Some peculiar, but physically significant relations between $\omega_1(t)$, $\omega_2(t)$, and $\gamma(t)$  
can make $\theta(t)$ time independent.  Here we consider instead the scenario where  
$\theta(t)$  changes with time. This is unavoidable if the process we want to implement 
implies boundary conditions for the parameters such that $\theta(0)\ne \theta(t_f)$, as in the examples below. 

{\it 2D Invariant.\label{2D}}
Urz\'ua et al. \cite{Urzua2019}, generalizing previous results in 1D  \cite{Guasti2002,Guasti2003}
and the work in \cite{Thylwe1998} for classical coupled oscillators, see also \cite{Castanos1994},
have recently found that 
the linear combination of operators (dots
stand for time derivatives hereafter)
\begin{equation}
     G(t)=u_1(t)  p_1 - \dot u_1(t)  q_1 + u_2(t)  p_2 - \dot u_2(t)  q_2,
     \label{invaG}
\end{equation}
satisfies the invariant equation $i  \partial  G / \partial t - [ H, G] = 0$, 
provided $u_1$ and  $u_2$ satisfy    
\beqa
\hspace*{-.3cm}\ddot{u}_{1}+\omega_{1}^2(t) u_{1} = \gamma(t)u_{2},\;\;\;\;\;
\ddot{u}_{2}+\omega_{2}^2(t) u_{2} = \gamma(t)u_{1},
\label{dyneqs}
\eeqa
{which are classical equations of motion  driven by a Hamiltonian (\ref{hamiltonianmw}). 
For any  state driven by $H(t)$, $\langle G(t)\rangle $ is the sum of two Wronskians $W_1[u_1(t),\langle q_1\rangle(t)]+W_2[u_2(t),\langle q_2\rangle(t)]$, where all functions in their arguments evolve
as Eq. (\ref{dyneqs}).   
The geometrical meaning of $W_i(t)$ is an ``oriented'' phase-space area formed by phase-space points $U_i(t)=\{u_i(0),\dot{u}_i(t)\}$,
$Q_i(t)=\{\langle q_i\rangle(t),\langle p_i\rangle(t)\}$ and the origin $\{0_i,0_i\}$. We consider two phase spaces, $i=1,2$, one for each oscillator.  
$W_i(t)$ is plus or minus the triangle area $A_i(t)$ depending on 
whether going from  $U_i$ to $Q_i$ needs an  
anticlockwise or clockwise displacement. 
For $\gamma=0$,  the two 
areas (and Wronskians) remain constant in time.  
When $\gamma\ne 0$ the individual Wronskians are not conserved. The conserved quantities are now $W_i(t)-\int_0^t \dot{W}_i(t') dt'= W_i(0)$, i.e., the initial phase-space oriented areas. The added terms cancel each other, namely, $\dot{W}_1=-\dot{W}_2=(u_1 \langle q_2\rangle-\langle q_1\rangle u_2)\gamma$, so that the sum  $W_1(t)+W_2(t)$
is the sum of oriented areas and it is   
constant.  This result is a particular case of the preservation of sums of oriented areas in classical Hamiltonian systems 
\cite{Arnold1989}.

We construct from ${G}$  a quadratic invariant that may become proportional to some relevant energy 
at boundary times by choosing specific boundary conditions for the $u_i$ and $\dot{u}_i$, $I=\frac 1 2  G^{\dagger} G$.
Designing  the $u_i$ we may manipulate the invariants and therefore the dynamics. From the $u_i$ we can as well get the Hamiltonian as demonstrated in the following two examples.      
{\it Controlled deflection.\label{deflection}}
A single particle is launched along a potential ``waveguide'' which is harmonic in the transversal direction.  Our goal is to deflect it, that is, manipulate the potential to change the waveguide direction, controlling 
the input/output scaling factor of the longitudinal velocity. 
To have waveguide potentials at the boundary times 
$t_b={0,t_f}$ we impose 
\begin{equation}
\label{gamma}
    \gamma(t_b) = \omega_1(t_b) \omega_2(t_b).
\end{equation}
As a consequence, $\Omega_l (t_b)=0$ and $\Omega_t(t_b) = [\omega_1^2(t_b) + \omega_2^2(t_b)]^{1/2}$. Thus, at boundary times, 
the potential is a harmonic ``waveguide'' with longitudinal direction defined by the       
angle
$
\theta(t_b)=\arctan [\omega_1(t_b)/\omega_2(t_b)].
$
The deflection angle $\Delta\theta=\theta(t_f)-\theta(0)$ can take any value between 0 and  $\pi/2$ for $\theta(t_f)\ge \theta(0)$.  
The condition (\ref{gamma}) in Eq. (\ref{dyneqs}) implies that 
$\ddot u_{1,2}(t_b)=0$, which also gives  
\beq
\label{boundaryrelation}
u_1(t_b) \omega_1 (t_b) = u_2(t_b) \omega_2 (t_b),
\eeq
i.e., the reference trajectories must start and end at $q_t(t_b)=0$, on the axis of the waveguide.
If the frequencies at $t_b$ are fixed, either $q_l(t_b)$, or one of the $u_i(t_b)$ can still be chosen freely.  

Rewriting the invariant $G$ in terms of the rotated variables $\{q_t, q_l\}$ and imposing $\dot u_{1,2}(t_b)=0$ we find that  
\beq
G(t_b)=
%
\frac{u_2(t_b)}{\sin{\theta(t_b)}}p_l,
\hspace{.8cm}
 I(t_b) =
\left[\frac{u_2(t_b)}{\sin\theta(t_b)}\right]^2\!\! \frac{p_l^2}{2},
\label{invatb}
\eeq
i.e., $I(t_b)$  is proportional to the longitudinal energy.

\begin{table}[t]
{\scriptsize
\begin{center}
\begin{ruledtabular}
\begin{tabular}{ccc}
&Initial waveguide&Final waveguide\\
\hline
\multirow{3}{*}{$\gamma$ const.}&
$\omega_1(0)$&$\omega_1(t_f)=\omega_2(0)$\\
&$\omega_2(0)$&$\omega_2(t_f)=\omega_1(0)$\\
&$\Omega_t(0)$&$\Omega_t(t_f)=\Omega_t(0)$\\
\hline
\multirow{3}{*}{$\omega_2$ const.}&
$\omega_1(0)$&$\omega_1(t_f)=\omega_2^2(0)/\omega_1(0)$\\
&$\omega_2(0)$&$\omega_2(t_f)=\omega_2(0)$\\
&$\Omega_t(0)$&$\Omega_t(t_f)=\frac{\omega_2(0)}{\omega_1(0)}\Omega_t(0)$\\
%
\end{tabular}
\end{ruledtabular}
\end{center}
}
\caption{Initial and final frequencies and angles defining the waveguides for $\gamma$-constant and $\omega_2$-constant protocols. The 
deflection angle $\Delta\theta=\theta(t_f)-\theta(0)$ determines the ratio $\omega_2(0)/\omega_1(0)$. \label{table1}}
\end{table}
\begin{figure}[t]
\begin{center}
\includegraphics[width=.5\textwidth]{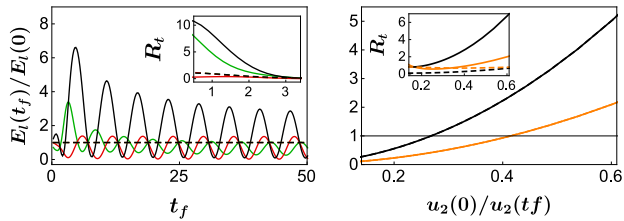}
\caption{Ratio of final to initial longitudinal energy for different process times $t_f$ (a) and for different scaling factors $u_2(0)/u_2(t_f)$ (b). The insets show the scaled transversal excitation $R_t=\Delta E_t/\Omega_t(t_f)$.
(a): initial longitudinal Gaussian wave packet  with $2^{1/2}\sigma=1$, $p_{l0} = 1$, and $q_{l0}= -4$ (green),  $q_{l0} = 0$ (red), and $q_{l0}=4$ (black). $\Delta\theta=\pi/4$ starting from $\omega_1(0) = 1$ and $\omega_2(0)=2.41$, 
using  linear ramps (solid lines) and an invariant-based protocol for $\gamma$ constant that produces $E_l(t_f)=E_l(0)$  (dashed lines). 
(b):  Initial longitudinal Gaussian wave-packet centered at the origin with $ p_{l0}= 1$ and $2^{1/2}\sigma=1$.  
$\Delta\theta=\pi/4$ with  $\omega_1(0)=1$ and $\omega_2(0)=2.41$ (orange curves), and $\Delta\theta=\pi/3$ with  $\omega_1(0)=1$ and $\omega_2(0)=3.73$ (black curves) for  constant-$\gamma$ processes (solid lines)
and constant-$\omega_2$ processes (dashed lines, overlapping with solid lines in main figure).
See Table \ref{table1} for values at $t=t_f$.  
\label{excitationsu2}}
\end{center}
\end{figure}
With Eq. (\ref{invatb}) we get
\beqa
\langle p_l(t_f)\rangle = F \langle p_l(0)\rangle,
\hspace{.8cm}
E_l(t_f)&=&F^2 E_l(0),
\label{elboundaries}
\eeqa
where $F=\frac{u_2(0)}{u_2(t_f)}\frac{\sin{\theta(t_f)}}{\sin{\theta(0)}}$ and   $E_l=\langle p_l^2/2\rangle$.
For some chosen deflection angle $\Delta \theta$ and waveguide frequencies $\Omega_t(t_b)$ 
we  may impose  any scaling factor by manipulating the ratio $u_2(0)/u_2(t_f)$. 
This scaling factor will affect all wave packets. Deflection angle,  velocity scaling, and waveguide 
compression/expansion factors can be chosen 
independently.

The Hamiltonian parameters are found inversely from Eq. \eqref{dyneqs}. We choose 
%
$
u_{1,2}=\sum_{k=0}^5 \alpha^{(1,2)}_k (t/t_f)^k,
$
%
with coefficients fixed so that $\dot u_{1,2}(t_b)=\ddot u_{1,2}(t_b)=0$, and the $u_{1,2}(t_b)$ are 
consistent with Eq. (\ref{boundaryrelation}).       

There are three external parameters, $\omega_1(t)$, $\omega_2(t)$ and $\gamma(t)$, but two coupled equations in Eq. \eqref{dyneqs}. 
Thus we may fix one of the external parameters or some combination.  
We consider two simple, not exhaustive,  possibilities: 
i) $\gamma$ constant, so initial and final $\Omega_t$ coincide;  and ii) $\omega_2$ constant, 
which implies a compression (transverse focusing useful to avoid transversal excitation) of the final waveguide with respect to the initial one, 
see Table \ref{table1}.  

\begin{figure}
\begin{center}
\includegraphics[width=.5\textwidth]{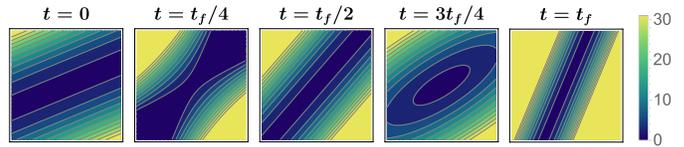}
\caption{ Snapshots of the top view of the 2D potential for  $E_l(t_f)=E_l(0)/2$ with constant $\omega_2$. $\omega_1(0)=1$ and $\omega_2=2.41$, deflection angle $\Delta\theta=\pi/4$ ($\omega_1(t_f)=2.41^2$) and process time $t_f=1$. The transversal frequency is compressed $2.41$ times, from $\Omega_t(0)=2.61$ to $\Omega_t(t_f)=6.29$, see Table \ref{table1}.\label{counterplot}}
\end{center}
\end{figure}

The initial state chosen for the numerical examples  is a product of  the ground state of the transversal harmonic oscillator and a minimum-uncertainty-product Gaussian  in the longitudinal direction centered at $q_{l0}$, with initial momentum $p_{l0}$,  
%
$
\psi_l(q_l, t=0)={[\sigma\sqrt{2\pi}]^{-1/2}}\,\,\,e^{i \, p_{l0} \,q_l}e^{-{(q_l-q_{l0})^2}/{(4\,\sigma^2)}}.
$
Firstly, we design a process that interchanges $\omega_1(t)$ and $\omega_2(t)$ with $\Delta\theta=\pi/4$ and constant $\gamma$, conceived to preserve the initial longitudinal velocities in the outgoing waveguide, $E_l(t_f)=E_l(0)$, and use linear ramps for the same boundary waveguides 
as a benchmark to compare the performance of the  invariant-based protocol.

Figure \ref{excitationsu2}a depicts  the final longitudinal energy.  For the linear ramps it oscillates with operation time. 
The envelope for the minima is at zero but the maximum tends for long  times to some value that depends on the initial wave packet. 
Contrast this with the full stability of the invariant-lead processes. They guarantee a fixed result, the final longitudinal energy being identical to the initial one for any initial wave packet. The transversal  excitation by the linear ramps in fast processes increases considerably as the  initial wave packet deviates  from the origin, while the transversal excitation 
in the invariant-based protocol is, in general, small and much more stable. It could be further suppressed by transverse focusing and/or optimizing  the  $u_i(t)$.  

Figure \ref{excitationsu2}b verifies that,  for some chosen deflection angle, we can scale the final longitudinal energy at will  in both scenarios ($\gamma$ or $\omega_2$ constant). 
Since the invariant does not affect the transversal direction, the transversal energy may be excited, but it depends  on the design of the $u_i(t)$ so it can be minimized or even suppressed.    
Figure \ref{counterplot}  provides 
snapshots of the evolution of the 2D potential for a $\omega_2$-constant processes  
that slows down the particle by a factor of two with deflection $\Delta\theta=\pi/4$. 

{\it State transfer.\label{sttr}}
Up to now we have considered real $u_j(t)$, but the coupled  Newton's equations 
admit purely real and purely imaginary solutions combined into  complex solutions. 
Exploiting this complex structure, $u_i=u_i^R+i u_i^I$, leads to interesting forms of the invariant. In particular the invariant may 
become proportional to the  uncoupled Hamiltonians at boundary times. 
Let us first drop the waveguide condition (\ref{gamma}) and go back to the laboratory frame variables $\{q_1, q_2\}$.
Defining annihilation operators in the usual manner, 
%
$
a_i(t)=\sqrt{{\omega_i(t)}/{2}}\,q_i+{i p_i}/{\sqrt{2\omega_i(t)}},\,\;i=1,2,
$
%
$G$ in Eq. (\ref{invaG}) may become $a_1$ or $a_2$ by certain choices of the $u_j$. Let us choose at initial time 
\begin{eqnarray}
u_1(0)={ic_0}/{\sqrt{2\omega_1(0)}},\;
\dot{u}_1(0)=-c_0\sqrt{{\omega_1(0)}/{2}},
\label{bch1n}
\end{eqnarray}
and $u_2(0)=\dot{u}_2(0)=0$ with $c_0$ real. This  implies  
%
$
G(0)=c_0a_1(0)$, and $I(0)={c_0^2}
a_1^\dagger(0) a_1(0)/2
$.
%
%
Instead, at final time we impose 
\begin{equation}
u_2(t_f)=ic_0/\sqrt{2\omega_2(t_f)},\;
\dot{u}_2(t_f)=-c_0\sqrt{\omega_2(t_f)/2},
\label{bch2n}
\end{equation}
together with $u_1(t_f)=\dot{u}_1(t_f)=0$,
so that 
$G(t_f)=c_0a_2(t_f)$, and $I(t_f)={c_0^2}
a_2^\dagger(t_f) a_2(t_f)/2$.
The same constant $c_0$ appears  in Eqs. (\ref{bch1n}) and (\ref{bch2n}) 
because the solutions of Eq. (\ref{dyneqs}) must satisfy 
$\frac{d}{dt}\{{\rm Im}[u_1^*(t)\dot{u}_1(t)+{u_2^*}(t)\dot{u}_2(t)]\}=0$ \cite{Thylwe1998}. 
The choice      
$c_0^2/2=\omega_1(0)$ 
gives 
$
I(0)=H_1(0)$
and $I(t_f)=[\omega_1(0)/\omega_2(t_f)]H_2(t_f)$, where we define the ``uncoupled Hamiltonians'' $H_j(t)\equiv \omega_j(t) a_j^\dagger(t) a_j(t)$. Eigenstates of $H_1(0)$ may thus be mapped into eigenstates of $H_2(t_f)$ by proper inverse engineering of 
the $u_j(t)$.
If $\omega_1(0)=\omega_2(t_f)$,  
\beq
\langle H_1(0)\rangle=\langle I(0)\rangle=\langle I(t_f)\rangle=\langle H_2(t_f)\rangle
\eeq
for all initial wavepackets. (Any other scale factor may be chosen.) 
 The system (\ref{dyneqs}), which now involves four real functions, $u_1^R(t), u_1^I(t), u_2^R(t), u_2^I(t)$, has to be solved inversely for $\omega_1(t), \omega_2(t)$ and $\gamma(t)$. 
The inversion is done following techniques developed for  trapped ions \cite{Palmero2017} or  mechanical systems \cite{Gonzalez-Resines2017},  
see appendices for a detailed account.
\begin{figure}[h]
\begin{center}
\includegraphics[width=.55\textwidth]{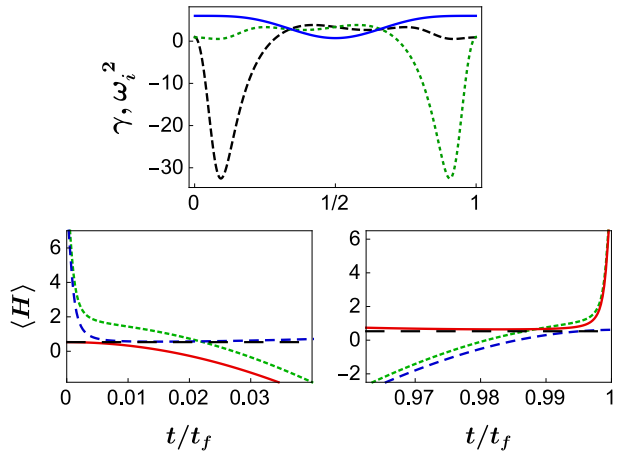}
\caption{Upper: Control parameters, $\omega_1^2$ (dashed black), $\omega_2^2$ (short-dashed green) and $\gamma$ (solid blue) vs. $t/t_f$, for an energy transfer from oscillator 1 to 2. Lower: $\langle H_1 \rangle$ in solid red, $\langle H_2 \rangle$ in dashed blue, $\langle H \rangle$ in short-dashed green, and $\langle I \rangle$ in long-dashed black, for initial (left) and final (right) parts of the process. $\omega_1(0)^2=\omega_2(t_f)^2=1$, $\omega_2(0)^2=\omega_1(t_f)^2=0.9$ and $\gamma(0)=\gamma(t_f)=6$;   $t_f=4$;  the system starts in a product state between the ground states of the uncoupled oscillators $H_1$ and $H_2$, not an eigenstate of the total Hamiltonian  \eqref{hamiltonianmw}.  \label{paramst} }
\end{center}
\end{figure}
Figure \eqref{paramst}a displays the resulting evolution of the control parameters for a specific example in which the frequencies $\omega_i$ swap their boundary values and $\gamma(0)=\gamma(t_f)$. Figure \eqref{paramst}b shows the expectation values of the total and the uncoupled Hamiltonians near the time boundaries, together with the constant expectation value of the invariant. 
Indeed $\langle H_2(t_f) \rangle=\langle H_1(0)\rangle$.

{\it Discussion.\label{discu}}
In some multidimensional systems with time-dependent control there are no point transformations that lead
to uncoupled normal modes. 
Our main point here is that in these ``coupled systems'', invariants of motion may still guide us to inversely 
design the time dependence of the controls for driving specific dynamics.

This  inversion procedure extends the domain of invariant-based engineering, which had been applied so far 
to one dimensional or uncoupled systems \cite{Guery2019}.
An important difference with respect to uncoupled systems is the diminished role of commutativity of Hamiltonian and invariant at boundary times. Commutativity, because of degeneracy, does not guarantee one-to-one mapping of eigenstates of the total Hamiltonian from initial to final configurations (see appendices).
One should then focus on the invariant itself for applications, and, if required,  rely on design freedom to keep other variables -e.g. the total energy- controlled. An alternative to be explored is to make use of a second invariant corresponding to a linearly independent 
set of classical solutions of Eq. (\ref{dyneqs}), $\{u_1'(t), u_2'(t)\}$, linearly independent with respect to $\{u_1(t), u_2(t)\}$ \cite{Thylwe1998}. Imposing boundary conditions to  the second set we would aim to control the second invariant as well, but the inversion problem becomes more demanding, as the
number of conditions double while the number of (common) controls remains the same.     

As for further open questions,  invariant-based engineering is known to be related to other STA approaches such
as counterdiabatic  driving for single oscillators \cite{Chen2011_062116}. It would  be of interest to connect the 
current work  with CD driving for coupled oscillators \cite{Duncan2018,Villazon2019}.  Finally, other boundary conditions on the $u_j$,  
see some examples in appendices, would allow to control other processes, different from the ones examined here. 
\acknowledgments{This work was supported by the Basque Country Government (Grant No. IT986-16), and 
by   PGC2018-101355-B-I00 (MCIU/AEI/FEDER,UE). 
E.T. acknowledges support from  PGC2018-094792-B-I00 (MCIU/AEI/FEDER,UE), CSIC Research Platform PTI-001, and CAM/FEDER  No. S2018/TCS- 4342 (QUITEMAD-CM).}
%

\clearpage
\appendix
\numberwithin{equation}{section}

\setcounter{equation}{0}
\setcounter{section}{1}
{{\large\bf{Appendices}}}
\\
\\for {\it Invariant-based inverse engineering of time-dependent, coupled harmonic oscillators}
by A. Tobalina et al. 
\section{A: Commutation of Hamiltonian and invariant at boundary times \label{boco}}

The eigenvectors of $I(t_b)$ in the waveguide deflection example are  highly degenerate, since a longitudinal plane wave multiplied by an 
arbitrary function of $q_t$ is a valid eigenvector with the same eigenvalue. 
This means that even if $I(t_b)$ commutes with $H(t_b)$ and shares some eigenvectors with $H(t_b)$ the vast majority of them are not eigenvectors of $H(t_b)$. 
This phenomenon -i.e., the existence of eigenvectors of one operator not shared with the other one- is well known but, since  it  sets an important difference with previous applications of invariant-based 
inverse engineering, we shall review briefly a few 
relevant aspects.  

 Let us consider a generic observable $A$ and an orthonormal set of eigenvectors of $A$ that forms a basis in the state space,
\beq
A | \phi_n^i\rangle = a_n |\phi_n^i\rangle; \hspace{1cm} \langle \phi_n^i | \phi_{n'}^{i'}\rangle = \delta_{n,n'}\delta_{i,i'},
\eeq
where $i=1,2...g_n$ is the index to distinguish the eigenstates in  the degenerate  subspace for 
eigenvalue $a_n$, and $g_n$ is the degree of degeneracy of $a_n$. 
Now let us introduce an operator $B$ that commutes with $A$. Since  $\langle \psi_1 | B | \psi_2\rangle = 0$ for any two eigenstates of $A$, $|\psi_1\rangle$ and $|\psi_2\rangle$, with different eigenvalues, we find  a block-diagonal matrix for $B$
in  the  $\{ |\phi_n^i\rangle\}$ basis, with blocks of dimension $g_n$ for each  eigenvalue, where any element within each block can be nonzero  \cite{Cohen-Tannoudji1991}. 

If $g_n =1$ for all $n$, that is, all the eigenvalues are non-degenerate, the matrix is diagonal as  all the blocks reduce to 
numbers $1\times 1$, and therefore the elements of the basis $\{ |\phi_n^i\rangle\}$ are eigenvectors of $B$. This applies in 
virtually all previous works on invariant-based shortcuts in 1D \cite{Chen2010a,Torrontegui2011,Tobalina2017}, in which the Hamiltonian and the invariant commute at initial and final time and are not degenerate, so they share the same eigenvectors at boundary times. Thus the inversely engineered protocol drives  an eigenstate of the invariant from an eigenstate of the initial Hamiltonian $H(0)$ to an eigenstate of the final Hamiltonian 
$H(t_f)$.  

If $g_n > 1$ for some $n$, the corresponding block does not reduce to a number and it is not, in general, diagonal. Thus, the elements of $\{ |u_n^i\rangle\}$ are not, in general, eigenvectors of $B$. The consequence for   
inverse engineering applications is that, if $I(t)$ is an invariant and $H(t)$ the Hamiltonian, an initial eigenvector 
of both $I(0)$ and $H(0)$ is guaranteed to be dynamically mapped at final time into an eigenvector of $I(t_b)$, but there is no guarantee that it will be simultaneously an eigenvector of $H(t_b)$.  

In particular, in the waveguide deflection example, the 
longitudinal energy may be 
conserved  
``asymptotically'' at the time boundaries\footnote{This terminology is borrowed from scattering theory. If a quantity is ``asymptotically conserved'' it has the same values before and after the interaction, but not necessarily during the process. In the current context there is no need to take infinite time limits, the conservation holds for times $t=0$ and $t_f$.}, but the process does not necessarily conserve the total energy.
A factorized  initial state with some longitudinal state multiplied by the transversal ground state 
will have at final time the same longitudinal energy that it had initially,  in a different direction,  but it can be  transversally excited.   
Avoiding only longitudinal excitations is  of interest \textit{per se}, but we may make use of the flexibility  of the shortcut design 
to minimize transversal excitation as well.
Similarly, in the state-transfer example the energy of oscillator 1 is transferred to oscillator 2, but oscillator 1 could be excited. 
%
%

In summary, commutativity of $H(t_b)$ and $I(t_b)$ plays a lesser role  in the 2D scenario, 
and may in fact be abandoned for different applications. In particular $H(t_b)$ and $I(t_b)$  do not commute in the state-transfer example. 
The following section (B) explores alternative boundary conditions for the $u_j(t_b)$
that imply different meanings for the invariant at the boundary times, and therefore different possible 
controlled processes.

\setcounter{equation}{0}
\setcounter{section}{2}
\section{B: Other boundary conditions.\label{obc}}
As the commutation of $I(t_b)$ with $H(t_b)$ does not guarantee the mapping among initial and final 
eigenstates of $H(t_b)$, we shall drop this condition and explore  other boundary conditions  
and forms of the invariant $I(t_b)$. 

For example, keeping the waveguide condition $\gamma(t_b) = \omega_1(t_b) \omega_2(t_b)$,
note the following alternative sets of  boundary  conditions and corresponding quadratic invariants:   
\begin{eqnarray}
\dot{u}_i(t_b)&=&0, u_1(t_b)\omega_2(t_b)=-u_2(t_b)\omega_1(t_b),
\nonumber\\  
I(t_b)&=&\frac{u_2^2(t_b)}{\cos^2\theta(t_b)}\frac{p_t^2}{2},
 \end{eqnarray}
where $i=1,2$ and the invariant at the boundary time $t_b$ is proportional to the transversal kinetic energy. 
As well, 
\begin{eqnarray}
{u}_i(t_b)&=&0,\; \dot{u}_1(t_b)\omega_1(t_b)=\dot{u}_2(t_b)\omega_2(t_b),
\nonumber\\  
I(t_b)&=&\frac{\dot{u}_2^2(t_b)}{\sin^2\theta(t_b)}\frac{q_l^2}{2};
 \end{eqnarray}
or 
\begin{eqnarray}
{u}_i(t_b)&=&0,\; \dot{u}_1(t_b)\omega_2(t_b)=-\dot{u}_2(t_b)\omega_1(t_b),
\nonumber\\  
I(t_b)&=&\frac{\dot{u}_2^2(t_b)}{\cos^2\theta(t_b)}\frac{q_t^2}{2},
 \end{eqnarray}
where the invariant at the boundary is proportional to the transversal potential energy. 

The boundary conditions imposed on the $u_i(t)$ and their derivatives do not need to be of the same type at $t=0$ and $t_f$. 
Designing $u_i(t)$ so as to satisfy at $t=0$ and $t_f$ different boundary conditions opens several 
control possibilities such as, for example, driving  the initial longitudinal energy into final transversal kinetic energy
or viceversa.

\setcounter{equation}{0}
\setcounter{section}{3}     
\setcounter{equation}{0}
\setcounter{section}{3} 
\section{C: Uncoupled limit (waveguide with no deflection) \label{hsa}}
In the main text the time-dependent guiding from an incoming to an outgoing waveguide implies formally time-dependent coupled
oscillators. If no deflection is desired, i.e. for $\Delta\theta=0$,  the easiest approach is to keep  the angle $\theta$ constant and 
the oscillators uncoupled, $\gamma=0$. We may thus identify $q_l=q_1$ and $q_t=q_2$, and the dynamics is separable into independent dynamical normal modes. There are linear invariants for each orthogonal direction, $G_i(t)=u_i(t)p_i-\dot{u}_i(t)q_i$, 
and corresponding quadratic invariants $I_i(t)=G_i^\dagger(t) G_i(t)/2$.  Focusing on 
the longitudinal direction, by imposing the boundary conditions 
\beqa
\omega_1(t_b)&=&0,
\nonumber\\ 
\dot{u}_1(t_b)&=&0,
\label{bc1D}
\eeqa
$G_1(t_b)=u_1(t_b)p_1$ is proportional to the longitudinal momentum. 
Consequences and applications, e.g. for cooling,  are worked out in \cite{Muga2020}. 

The transversal direction evolves independently. The simplest possibility  is a waveguide with $\omega_2$ constant. Of course, compressions and or expansions may as well be designed for $\omega_2(t_f)\ne \omega_2(0)$ via invariants, 
free from any transversal excitation as is well known  \cite{Chen2010_063002}.   
In the current formal framework, making use of complex solutions as in the main text  we would impose   
\begin{eqnarray}
u_2(0)&=&\frac{ic_0}{\sqrt{2\omega_2(0)}},
\;\;\;\,\dot{u}_2(0)=-c_0\sqrt{\frac{\omega_2(0)}{{2}}},
\nonumber\\
u_2(t_f)&=&\frac{ic_0}{\sqrt{2\omega_2(t_f)}},
\;\dot{u}_2(t_f)=-c_0\sqrt{\frac{\omega_2(t_f)}{{2}}},
\label{bctr}
\end{eqnarray}
with $c_0$ real.     
This choice of boundary conditions implies  
\begin{eqnarray}
G_2(0)&=&c_0a_2(0), 
\nonumber\\
I_2(0)&=&\frac{1}{2}G_2^\dagger(0) G_2(0)=\frac{c_0^2}{2}
a_2^\dagger(0) a_2(0),
\end{eqnarray}
and at final time
\begin{eqnarray}
G_2(t_f)&=&c_0a_2(t_f), 
\nonumber\\
I_2(t_f)&=&\frac{1}{2}G^\dagger(t_f) G(t_f)=\frac{c_0^2}{2}
a_2^\dagger(t_f) a_2(t_f).
\label{bctr2}
\end{eqnarray}
Choosing   
$c_0^2/2=\omega_2(0)$  gives $I_2(0)=H_2(0)$ and $I_2(t_f)=\frac{\omega_2(0)}{\omega_2(t_f)}H_2(t_f)$, where 
$H_2(t)=\omega_2(t)a_2^\dagger(t)a(t)$. 
Eigenstates of $H_2(0)$ may thus be mapped into eigenstates of $H_2(t_f)$ by proper inverse engineering of 
$u_2(t)$. The uncoupled equation for the transverse oscillator 2 in Eq. (7)
which now involves two real functions, $u_2^R(t), u_2^I(t)$, has to be solved inversely for $\omega_2(t)$. 

This approach is equivalent to the usual one using the Ermakov equation \cite{Chen2010_063002}.
By writing $u_2(t)$ in polar form, $u_2(t)=\rho_2(t)e^{i\phi_2(t)}$, the harmonic oscillator equation splits into two coupled equations. One 
of them gives $\rho_2^2(t)\dot{\phi}_2=K$, with $K$ constant, and the other one is the Ermakov equation for $\rho_2(t)$ \cite{Thylwe1998,Guasti2002,Guasti2003},
\beq
\ddot{\rho}_2+\omega_2^2(t)\rho_2=\frac{K^2}{\rho_2^3}.
\eeq
The invariant $I_2(t)$ becomes
\beqa
I_2&=&I_{LR}-\frac{K}{2},
\nonumber\\
I_{LR}&=&\frac{1}{2}\left[\left(\frac{Kq_2}{\rho_2}\right)^2+\left(\rho_2 p_2-\dot{\rho}_2 q_2\right)^2\right],
\eeqa
where $I_{LR}$ is the ``Lewis-Riesenfeld'' invariant \cite{Lewis1969}.
With the usual choice $K=\omega_2(0)$, 
and boundary conditions 
\beqa
\rho_2(0)&=&1,\,\dot{\rho}_2(0)=\ddot{\rho}_2(0)=0,
\nonumber\\
\rho_2(t_f)&=&\left(\frac{\omega_2(0)}{\omega_2(t_f)}\right)^{1/2},\,\dot{\rho}_2(t_f)=\ddot{\rho}_2(t_f)=0,
\eeqa
which are equivalent to Eq. (\ref{bctr}) for $c_0=[2\omega_2(0)]^{1/2}$,   
$I_{LR}(0)=\omega_2(0)[a_2^\dagger(0) a_2(0)+1/2]$, whereas 
$I_{LR}(t_f)=\omega_2(0)[a_2^\dagger(t_f) a_2(t_f)+1/2]$.

\setcounter{equation}{0}
\setcounter{section}{4} 
\section{D: Inverse engineering of a state transfer\label{designst}}

Here we present the details of the inverse engineering for the state transfer protocol, we follow a method similar to the ones in refs. \cite{Palmero2017,Gonzalez-Resines2017}. 
Assuming that the values of the control parameters at boundary times are set, we start by designing a $\gamma(t)$ that satisfies the boundary values $\gamma(t_b)$ and that has zero first and second derivatives at the boundaries for smoothness. We use a sum of cosines ansatz, 
\beq
\gamma(t)=\sum_{k=0}^4 a_k \cos \left(\frac{k \, \pi t}{t_f}\right),
\eeq
which meets the boundary conditions with just five terms. 
The coefficient $a_4$ is left free for now. Then we design the imaginary part of the dynamics, again using sums of cosines,
\beqa
u_1^I(t)=\sum_{i=0}^6 b_i \cos \left(\frac{i\, \pi \, t}{t_f}\right),
\nonumber\\
 u_2^I(t)=\sum_{j=0}^6 c_j \cos \left(\frac{j\, \pi \, t}{t_f}\right).
\eeqa
Coefficients $\{b,c\}_{1-5}$ are fixed so that the real reference trajectories satisfy the boundary conditions 
for $u_{1,2}(t_b)$ and its derivatives, and so that the frequencies $\omega_i(t)$ have the desired boundary values, which amounts to satisfying 
\beqa
&\ddot u&_1^I (0)= - \omega_1(0)^2,
\hspace{.5cm}
\ddot u_1^I (t_f)=\gamma(t_f) \sqrt{\frac{\omega_1(0)}{\omega_2(t_f)}},\nonumber\\
&\ddot u&_2^I(0)= \gamma(0),
\hspace{.3cm}
\ddot u_2^I(t_f)=\omega_2(tf) \sqrt{\omega_1(0)\omega_2(t_f)}.
\label{condim}
\eeqa
Note, from the expression of the frequencies 
\beq
\omega_{1,2}(t)^2=\frac{\gamma(t) u_{2,1}^I(t) - \ddot u_{1,2}^I(t)}{u_{1,2}^I(t)},
\eeq
that, even if the conditions in Eq. \eqref{condim} are fulfilled, we may encounter indeterminacies at boundary times (some $u_{1,2}(t_b)$ become $0$). Thus, we have to impose additional boundary conditions for consistency using L'Hopital's rule, 
\beqa
 u_1^{I\,(3)}(t_f)&=&0,\nonumber\\
 u_1^{I\,(4)}(t_f)&=& -\gamma(t_f)  \sqrt{\frac{\omega_1(0)}{\omega_2(t_f)}}\left[\omega_1(t_f)^2 + \omega_2(t_f)^2\right],\nonumber\\
 u_2^{I\,(3)}(0)&=&0,\nonumber\\
 u_2^{I\,(4)}(0)&=& -\gamma(0) \left[\omega_1(0)^2 + \omega_2(0)^2 \right].
 \eeqa
Coefficients $\{b,c\}_6$ are left yet undetermined. In the next step, we numerically solve the real equations of motion with the already designed control parameters for the initial conditions  and  find with an optimization subroutine the value of the coefficients 
that have been left free to satisfy  the final boundary conditions. For the specific example presented in the main text the ``free'' coefficients take the values  $a_4=-0.659$, $b_6=-0.383$ and $c_6=-0.383$.

\bibliography{Bibliographyb}{}
\bibliographystyle{sofia}

\end{document}